\documentclass[a4paper,11pt]{article}
\usepackage[a4paper,left=0.99in, right=0.99in,top=1.2in, bottom=1.2in]{geometry}

\usepackage{amsmath}
\usepackage{amssymb}
\usepackage{fixmath}
\usepackage{hyperref}
\usepackage{color}
\usepackage{cite}
\usepackage{graphicx}
\usepackage{xspace}

\definecolor{red}{rgb}{1,0,0}

\def\be{\begin{equation}}
\def\ee{\end{equation}}
\numberwithin{equation}{section}
\newcommand{\jet}{\text{jet}}
\newcommand{\eg}{\emph{e.g.}\xspace}
\newcommand{\ie}{\emph{i.e.}\xspace}
\newcommand{\pythia}{\textsc{Pythia}\xspace}
\newcommand{\herwig}{\textsc{Herwig}\xspace}
\newcommand{\forward}{\textsc{forward}\xspace}
\newcommand{\as}{{\alpha_s}}
\newcommand{\GeV}{\text{GeV}}
\newcommand\vhlib{{\sc AVHLIB }}

\title{
\vskip 1cm
Single and double inclusive forward jet production\\ 
at the LHC at $\sqrt{s} = 7$ and $13\,$ TeV
\vskip 1cm
}
\author{
Marcin Bury$^1$, Michal Deak$^1$, Krzysztof Kutak$^1$ and  
Sebastian Sapeta$^{1,2}$ \\ \\
$^1$ {\small\it The H.\ Niewodnicza\'nski Institute of Nuclear Physics PAN,}\\ 
{\small\it Radzikowskiego 152, 31-342 Krak\'ow, Poland}\\\\
$^2$ {\small\it 
Theoretical Physics Department, CERN, Geneva, Switzerland }\\
}

\date{}

\begin{document}
\maketitle

\vspace{-28em}
\begin{flushright}
  IFJPAN-IV-2016-6 \\
  CERN-TH-2016-077 \\
\end{flushright}
\vspace{14em}

\vspace{10em}
\begin{abstract}
We provide a description of the transverse momentum spectrum of single inclusive
forward jets produced at the LHC, at the center-of-mass energies of 7 and 13
TeV, using the high energy factorization~(HEF) framework. 
We subsequently study double inclusive forward jet production and, 
in particular, we calculate contributions to azimuthal angle distributions
coming from double parton scattering.
We also compare our results
for double inclusive jet production to those obtained with the \pythia Monte Carlo
generator. This comparison confirms that
the HEF resummation acts like an initial state parton shower. 
It also points towards the need to include final state 
radiation effects in the HEF formalism.  
\end{abstract}

\section{Introduction}

Processes with jets produced at forward rapidities offer unique access to the
corner of phase space where the magnitude of the longitudinal momentum of one of
the incoming partons is close to
that of the proton, whereas the other parton carries very small fraction of
proton's longitudinal momentum, $x \ll 1$. 
The latter leads to appearance of large logarithms, $\alpha_s\ln(1/x)$, from initial
state emissions, which should be resummed, \eg by means of the BFKL equation
\cite{Kuraev:1977fs,Balitsky:1978ic,Kuraev:1976ge}, for moderate values of $x$,
or its nonlinear extensions
\cite{Balitsky:1995ub,JalilianMarian:1997dw,Kovchegov:1999yj,Kovchegov:1999ua,Kovner:2005nq,Kutak:2011fu} if the $x$
is small.
The resummation leads to gluon distributions that depend not only on $x$, but
also on the transverse component of gluon's four-momentum, $k_t$, and  the
hadronic cross section factorizes into a convolution of such unintegrated gluon
distributions and the corresponding off-shell matrix elements.
This approach, commonly referred to as \emph{$k_t$-factorization} or 
\emph{high energy factorization}~(HEF) \cite{Catani:1990eg}, will be the
basic framework used to study forward jet production in this work.
It is worth mentioning that HEF-based approaches to forward jet physics
stimulated very interesting recent theoretical developments like the effective
TMD approach to dilute-dense collisions \cite{Kotko:2015ura} or dedicated
applications of soft-collinear effective theory~(SCET) formalism~\cite{Rothstein:2016bsq}.
 
Alternatively to the above, one can attempt to calculate predictions for the
production of forward jets using general purpose Monte Carlo~(MC) programs,
such as \pythia~\cite{Sjostrand:2014zea} or \herwig~\cite{Bellm:2015jjp}, which
are based on the collinear factorization of a $2\to 2$ process, supplemented
with an initial- and a final-state parton shower~(PS).
The advantage of this approach is that it allows one to include a range of
potentially important physical effects, such as multi-parton interactions,
final-state radiation and non-perturbative corrections. 
At the same time,
however, it lacks formal resummation of terms enhanced with $\alpha_s\ln(1/x)$ and the
correct behaviour at low $x$ is only modelled by appropriate initial condition
for evolution of the collinear parton density functions.

The collinear-factorization MC tools employing collinear factorization are
currently very well developed and have also been successfully used to describe
production of forward jets at the LHC (for the latest review
see~\cite{Sapeta:2015gee}). 
On the other hand, the recently  developed HEF-based tools like
\vhlib~\cite{Bury:2015dla} and {\sc LxJet}~\cite{Kotko_LxJet} form a milestone
in HEF calculations, they are, however, still at the stage where they can profit
from further improvements, as they do not include effects of
multi-parton scattering nor a modelling of the final state interactions as for instance is the case in HEF Monte Carlo generator CASCADE \cite{Jung:2010si}.
 
The aim of this work is to make a few steps towards more realistic theoretical
description of the forward jet production in the framework of the high energy
factorization by investigating the importance of a range of physical effects
neglected in earlier analyses. These include: contributions to the hard
scattering coming from diagrams with off-shell quarks, contributions from
double-parton scattering and the effects of final-state radiation.

The article is organized as follows. 
We start in Section~\ref{sec:single-inclusive} from studying how the current
state-of-the-art HEF framework fares in description of the recent LHC data for
the single inclusive forward jet production. 
Then, in Section~\ref{sec:dijet}, we turn to dijet production and study
potential importance of several physical components that were not considered in
the description of forward jet production so far~\cite{vanHameren:2014lna,
vanHameren:2014ala}.
In particular, in Section \ref{sec:dijet-hef}, we use the results from single
inclusive jet production to construct double-parton scattering~(DPS)
contributions to dijet processes and assess their relevance. 
Then, in Section \ref{sec:pythia}, we compare HEF results for forward dijet
spectra with those from the \pythia MC generator, which include full parton
shower, \ie the initial-~(ISR) and the final state radiation~(FSR).
All the above allows us to quantify effects that are
currently not included in phenomenological analyses within the HEF framework.

\section{Single inclusive forward jet production}
\label{sec:single-inclusive}

The single inclusive jet production is a process which can directly probe
partonic content of the proton without a need for large corrections from
fragmentation functions. 
What makes it interesting is the possibility
to apply the appropriate formula already at leading order in high energy
factorization. This is to be contrasted with collinear factorization, where
the $2\to 1$ emission vertex vanishes identically and one has to account for
higher order corrections either at fixed order of $\as$, or with a parton
shower. 
 
The single inclusive jet production process can be schematically written as 
\begin{equation}
  A + B \mapsto a + b \to \text{jet} + X\,,
  \label{eq:single-incl-def}
\end{equation}
where $A$ and $B$ are the colliding hadrons, each of which provides a parton,
respectively $a$ and $b$. $X$ corresponds to undetected, real radiation and the
beam remnants from the hadrons $A$ and $B$ are understood in the above equation.

The longitudinal kinematic variables read
\begin{equation}
  x_1 = \frac{1}{\sqrt{s}}\, p_{t,\jet}\, e^{y_\jet}\,,
  \qquad \qquad \qquad
  x_2 = \frac{1}{\sqrt{s}}\, p_{t,\jet}\, e^{-y_\jet}\,,
\end{equation}
where $s = (p_A+p_B)^2$ is the total squared energy of the colliding hadrons
while $y_\jet$ and $p_{\jet}$  are the rapidity and transverse momentum of the
leading final state jet, respectively.

The hybrid, high energy factorization formula for the
process~(\ref{eq:single-incl-def}), justified for configurations with $x_1\gg
x_2$, at the leading $\ln\left(1/x\right)$ accuracy\footnote{At the NLO level, there exists an extension of this formula for single particle production \cite{Chirilli:2011km,Altinoluk:2014eka}.}, reads~\cite{Dumitru:2005gt}
\begin{equation}
  \frac{d\sigma}{dy_\jet dp_{t, \jet}} 
   = \frac{1}{2}
  \frac{\pi\, p_{t, \jet}}{(x_1x_2 s)^2}
  \sum_{a,b,c}
  \overline{|{\cal M}_{ab^*\to c}|}^2
   x_1f_{a/A}(x_1,\mu^2)\, {\cal F}_{b/B}(x_2,p_{t, \jet}^2,\mu^2)\,,
   \label{eq:int-phi}
\end{equation}
where ${\cal F}$ is a generic notation for the transverse momentum
dependent parton density~(TMD), which is a  function of the longitudinal
momentum fraction $x$, transverse momentum $k_t$, and the factorization scale 
$\mu$. 
 
TMDs can be obtained in several ways. In particular, they
can be constructed from collinear parton densities via the KMR procedure
\cite{Kimber:1999xc,Kimber:2001sc} or, in the approximation in which the gluon
dominates over quarks, they can be obtained as solutions of low-$x$ evolution equations \cite{Lublinsky:2001yi,GolecBiernat:2001if,Kutak:2003bd,Kutak:2004ym,Albacete:2010sy,Lappi:2015fma}. 
The function $x_1f_{a/A}(x_1,\mu^2)$ is a generic expression for the collinear
parton density.  The matrix elements  $|{\cal M}_{ab^*\to c}|^2$ can be obtained
via application of the helicity-based formalism
\cite{vanHameren:2012uj,vanHameren:2012if,vanHameren:2013csa} for off-shell
partons or the \emph{parton reggezation approach}~\cite{Nefedov:2013ywa}. 
The following channels contribute to the single jet production in HEF approach
\begin{equation}
  gg^* \to g\,,
  \qquad \quad
  qg^* \to q\,,
  \qquad \quad
  gq^* \to q\,,
  \qquad \quad
  \bar{q}q^* \to g\,.
\end{equation}
The
explicit expressions for the corresponding matrix elements are collected in 
appendix~\ref{app:single-inclusive-matrix-elements}.  

We now turn to predictions for the transverse momentum spectra of the  single
inclusive forward jets at the LHC. We have performed our calculations at the
center-of-mass energies of $\sqrt{s} = 7$ and 13 TeV. 
The event selection was applied by requiring a leading jet with 
$p_{t, \jet} > 35\, \GeV$ in the rapidity window of $3.2<|y_\jet|<4.7$,
following the cuts used in the CMS analyses of Refs.~\cite{Chatrchyan:2012gwa,CMS13}.
For the on-shell partons, denoted by $x_1f_{a/A}(x_1,\mu^2)$ in
Eq.~(\ref{eq:int-phi}), we used the distribution from the CT10 NLO
set~\cite{Lai:2010vv}.
For the off-shell partons, ${\cal F}_{b/B}(x_2,p_{t, \jet}^2,\mu^2)$, we chose
the following set of distributions:
\begin{itemize}
  \item
  The ``KS nonlinear'' unintegrated gluon density~\cite{Kutak:2012rf}, which
  comes from an extension of the BK equation~\cite{Kutak:2003bd,Kutak:2004ym}
  following the prescription of Ref.~\cite{Kwiecinski:1997ee} to include
  kinematic constraint on the gluons in the chain, non-singular pieces of the
  splitting functions, as well as contributions from sea quarks. The parameters
  of the gluon were set by the fit to $F_2$ data from HERA. 
  \item
  The ``KS linear'' gluon~\cite{Kutak:2012rf}, determined from linearized
  version of the equation described above.
  \item
  The ``KShardscale nonlinear''  unintegrated gluon density \cite{Kutak:2014wga}
  obtained from the ``KS nonlinear'' gluon by performing Sudakov
  resummation of soft emissions between scales given by a hard probe and the
  scale defined by emission of gluons from the gluonic chain.
\item
  The ``KShardscale linear'', determined from linearized
  version of the equation described above.
\item
  The ``DLC2016'' (Double Log Coherence) \cite{Kutak:2016mik} unintegrated
parton densities for the gluon and the quarks, determined following the KMR
prescription \cite{Kimber:1999xc,Kimber:2001sc}.  These distributions are
obtained from the standard collinear PDFs supplemented with angular ordering
imposed at the last step of evolution and resummation of soft emissions. The
Sudakov form factor ensures no emissions between the scale of the gluon
transverse momentum, $k_t$, and the scale of the hard process, $\mu$. The upper
cutoff in the Sudakov form factor is chosen such that it imposes angular
ordering in the last step of the evolution. The unintegrated parton
distributions used in our study are based on CT10 NLO \cite{Lai:2010vv}.
\end{itemize}

All HEF predictions in this and in the following section were obtained with the
\forward program~\cite{forward}. The code implements the hybrid high energy
factorization for the single and double jet production and it is capable of
using both gluon and quark off-shell parton distributions.

As we see from the above list, all parametrizations, except that of DLC2016,
provide only off-shell gluons and neglect off-shell quarks by assuming that
their relative contribution is much smaller.
The DLC2016 distributions provide the full set of partons and this gives us
unique opportunity to verify this assumption.

\begin{figure}[t]
\centering
  \includegraphics[width=0.48\textwidth]{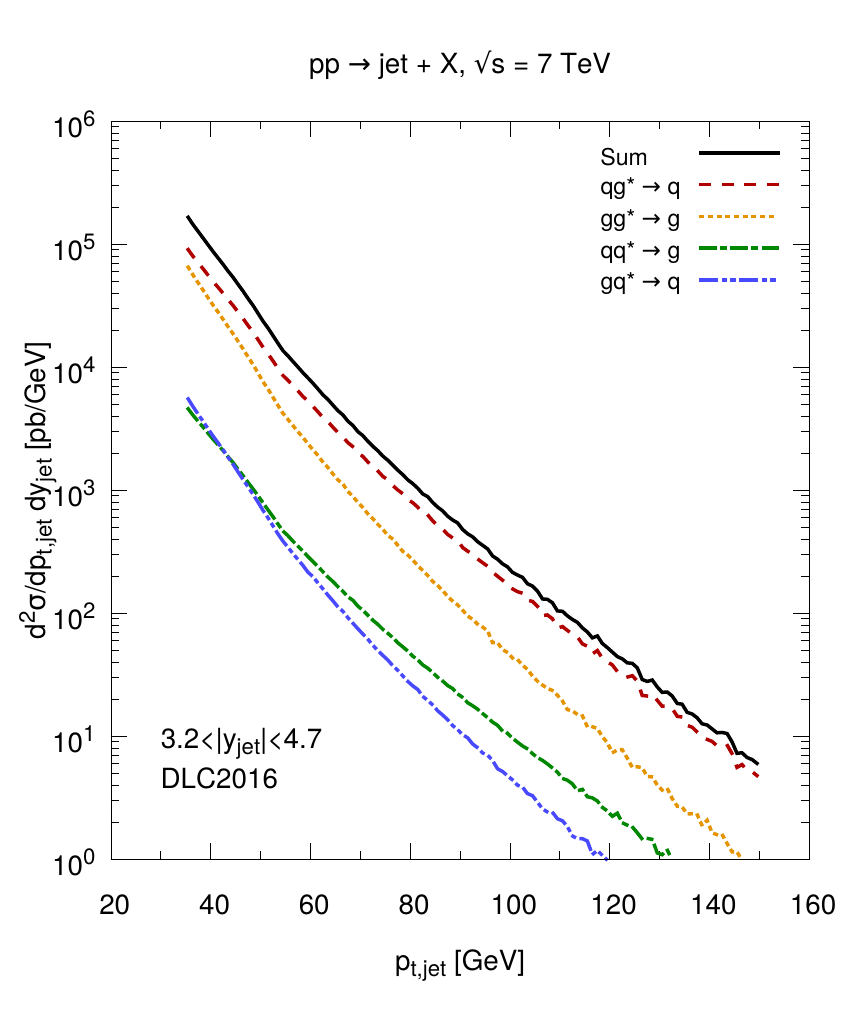}
  \includegraphics[width=0.48\textwidth]{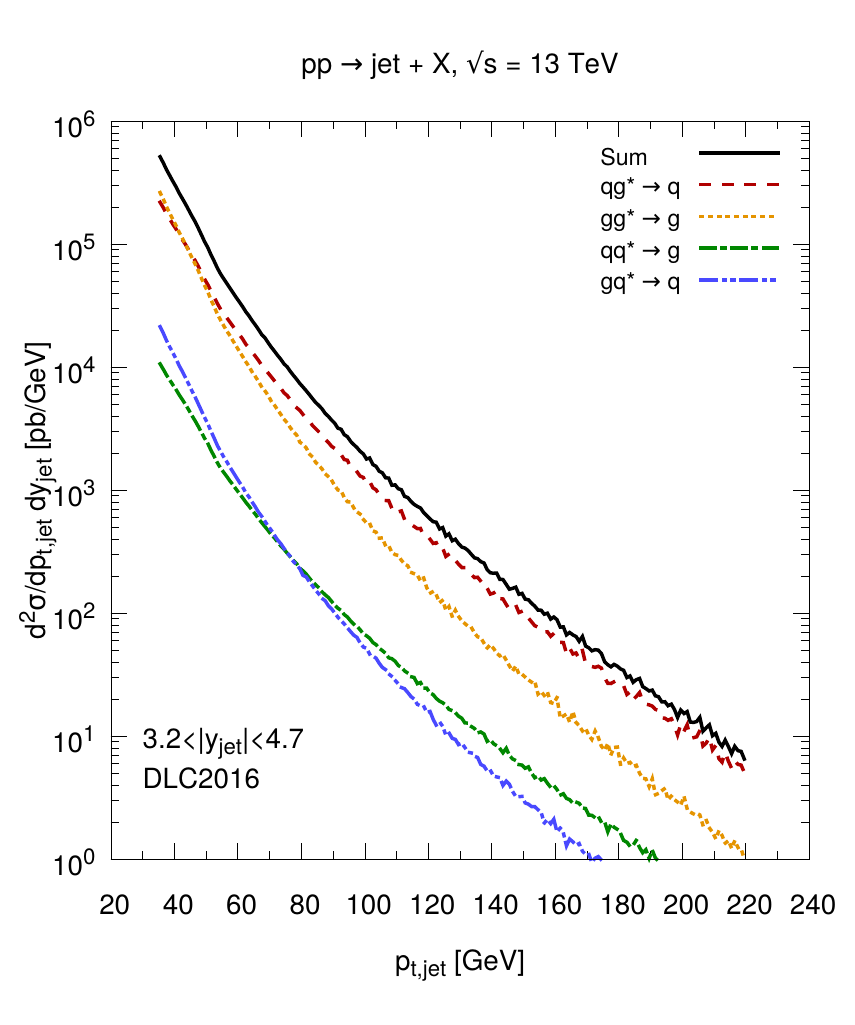}
  
  \caption{\small{
  Single inclusive forward jet production.  Comparison between different
channels contributing to the spectrum of jet's transverse momentum. The results
use DLC2016~\cite{Kutak:2016mik} parametrization for the off-shell partons.} 
  }
  \label{fig:single-incl-q1}
\end{figure}

In Fig. \ref{fig:single-incl-q1} we show predictions for various contributions
to the single inclusive jet production obtained using Eq.~(\ref{eq:int-phi})
with the DLC2016 off-shell partons.
It is evident that
the off-shell quark contribution can be indeed effectively neglected and we
can proceed just with the off-shell gluons in the initial state.
The second interesting observation is that 
the $qg^*\rightarrow q$ channel gives larger contribution than
$gg^*\rightarrow g$ at high transverse momentum while the two channels
contribute comparably at lower $p_{t,\jet}$.

\begin{figure}[t]
\centering
  \includegraphics[width=0.48\textwidth]{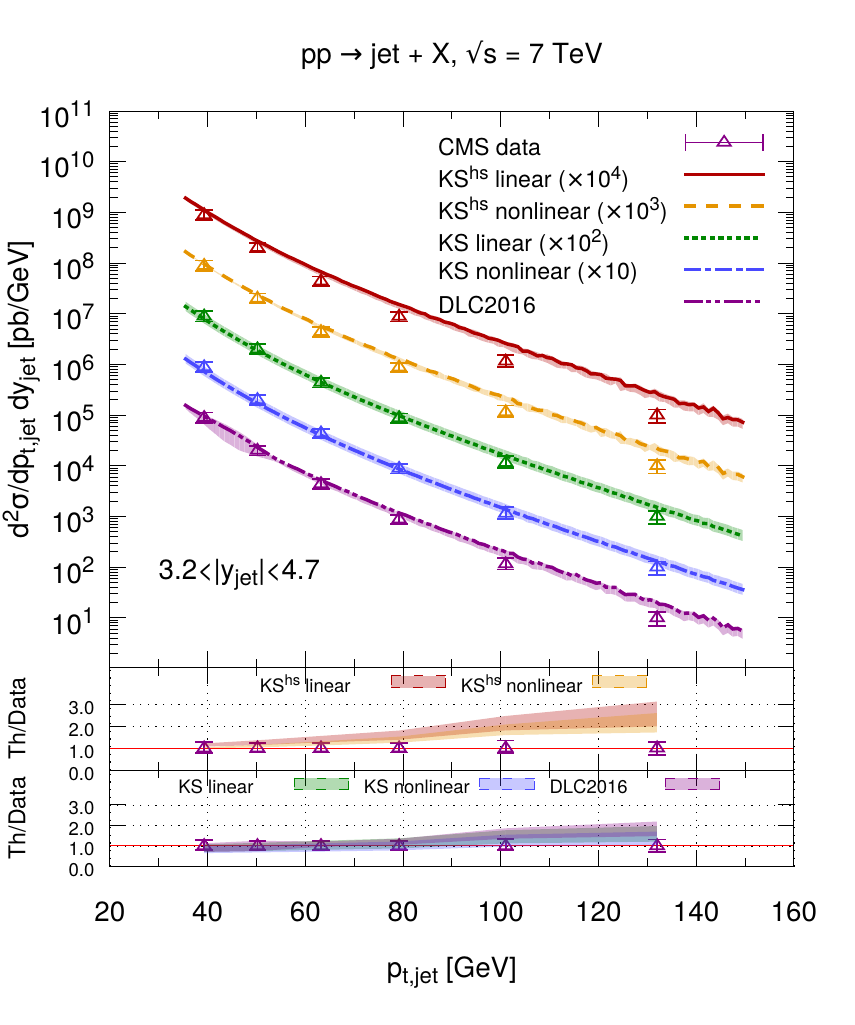}
  \includegraphics[width=0.48\textwidth]{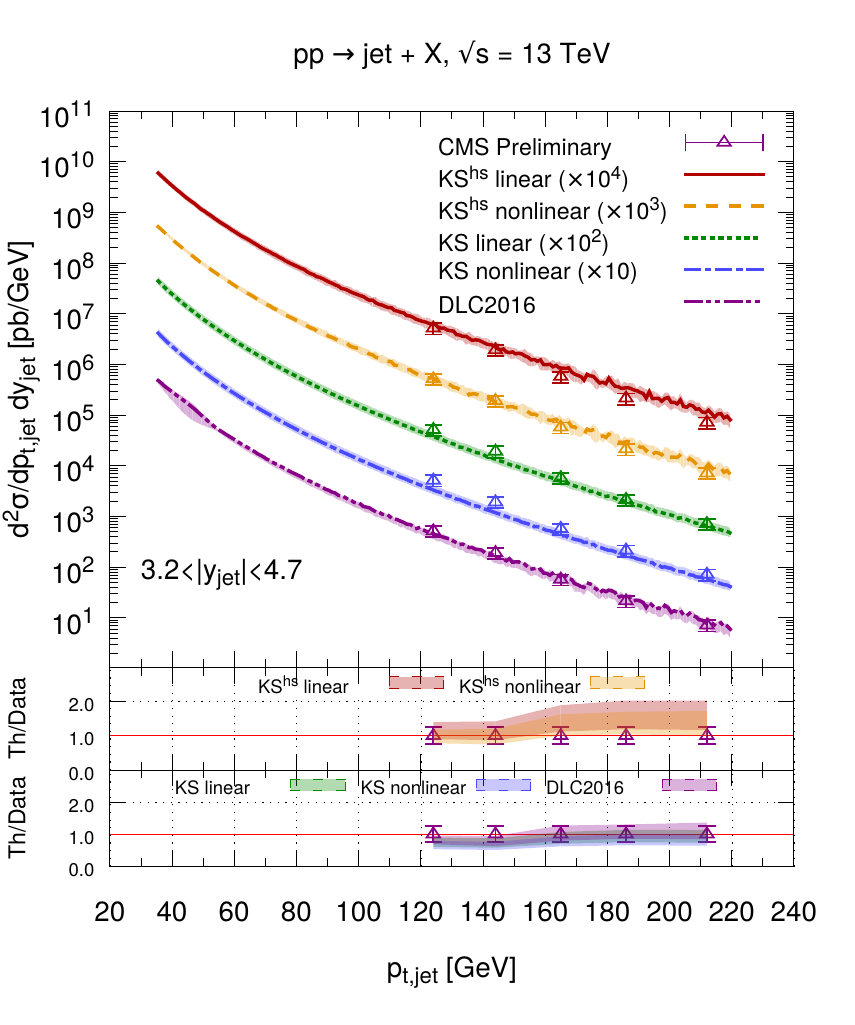} 
  
  \caption{\small{
    Single inclusive forward jet production. Comparison of predictions for the
    transverse momentum distributions of a jet with CMS data at
    7~\cite{Chatrchyan:2012gwa} and 13~TeV~\cite{CMS13}.
    The bands correspond to different gluon distributions used for calculations.
    The width of the bands comes from varying the factorization and 
    renormalization scales by factors
    $\frac{1}{2}$ and $2$ around the central value equal to
    $\mu_F=\mu_R=p_{t,\jet}$. In case of KShardscale for 13 TeV, because of
    limited size of the available grid, we could vary the scale only by the
    factors $\frac12$ and $\frac32$.
    For better visibility, data and predictions with various unintegrated gluons
    were multiplied by factors $10^n$, with $n=0,\ldots,4$.
    }
  }
  \label{fig:single-incl-TMDs}
\end{figure}

As the off-shell quark contributions are negligible, it is justified to use all
of the off-shell gluon sets listed above as an input for predictions of single
inclusive jet spectra.
The corresponding results are shown
in Fig. \ref{fig:single-incl-TMDs}, where the upper panel shows the absolute
distributions, whereas the two lower panels show theory-to-data ratio. 
We observe good compatibility of the predictions and the 7 and 13-TeV CMS
data~\cite{Chatrchyan:2012gwa, CMS13} across a range of unintegrated gluon
distributions.
We believe that this is a consequence of the TMD factorization applied to
low-$x$ physics \cite{Dominguez:2011wm}, which states that the same gluon
density (if saturation effects are negligible in the considered phase space
region) is to be used for the $F_2$ structure function and for the single
inclusive gluon production.

\section{Forward dijet production}
\label{sec:dijet}

Dijets can be produced either in a \emph{single-parton scattering}~(SPS)
\begin{equation}
  A + B \mapsto a + b \to \text{jet} + \text{jet} + X\,,
\end{equation}
where the partons $a$ and $b$ interact through a $2\to 2$ process, or in a
\emph{double parton scattering}~(DPS)
\begin{equation}
  A + B \mapsto a_1 + b_1 + a_2 + b_2 \to \text{jet} + \text{jet} + X\,,
\end{equation}
in which each of the incoming hadrons provides two-parton pairs $a_i+b_i$, which
in turn undergo two $2\to 1$ scatterings. 
The DPS can be thought of as the single inclusive jet production process of
Eq.~(\ref{eq:single-incl-def}) squared.

\subsection{Results within HEF formalism}
\label{sec:dijet-hef}

In general, in order to comply with the state of the art of theoretical
development, description of the SPS process needs corrections from the improved
TMD factorization~\cite{Kotko:2015ura}, as it gets contribution from the so-called quadrupole configurations of colour glass condensate~(CGC) states and the
latter are important in the non-linear domain.

In the present letter, however, we focus on the region of azimuthal distance
between the two leading jets, $\Delta\phi$, where the bulk of linear and
nonlinear KS densities \cite{Kutak:2012rf} give comparable
results~\cite{vanHameren:2014lna} and our aim is just to quantify the potential
corrections coming from other physical effects like DPS contributions and 
final-state parton shower.
Encouraged by the good description of the single inclusive jet production,
presented in Section~\ref{sec:single-inclusive}, we aim at evaluation of the DPS
contribution to inclusive dijet production in order to assess its relative
impact with respect to~SPS.

\begin{figure}[p]
\centering
  \includegraphics[width=0.48\textwidth]{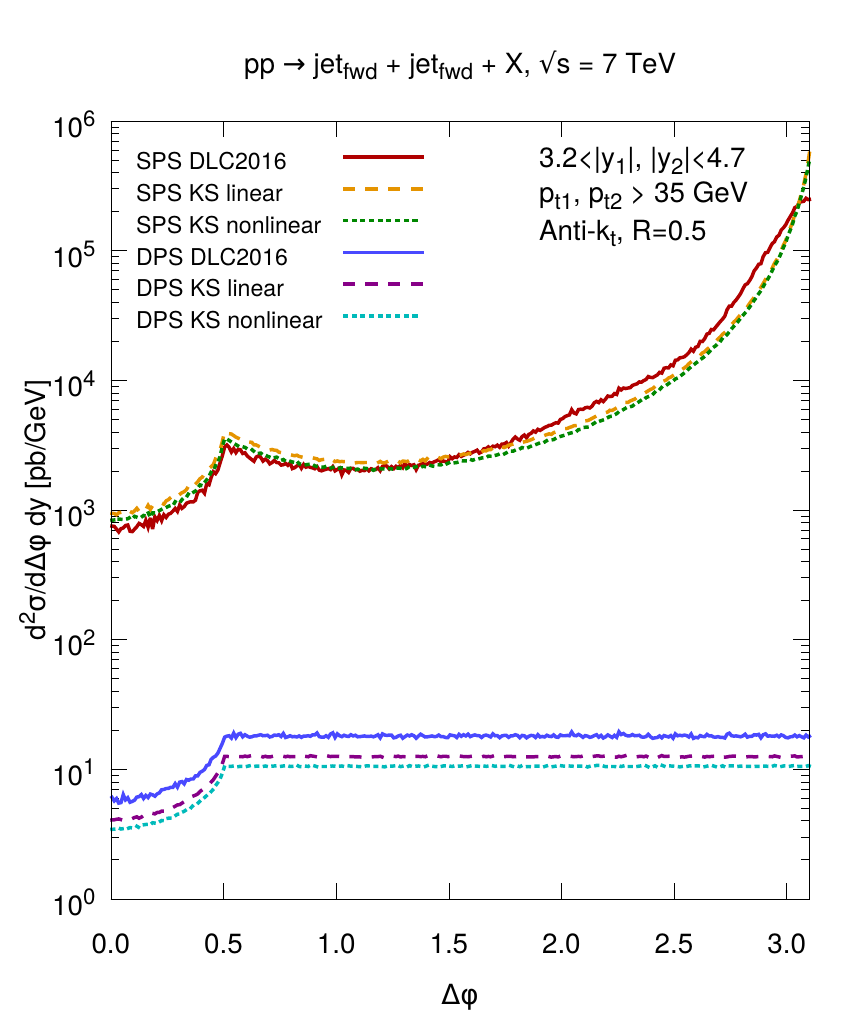}
  \hfill
  \includegraphics[width=0.48\textwidth]{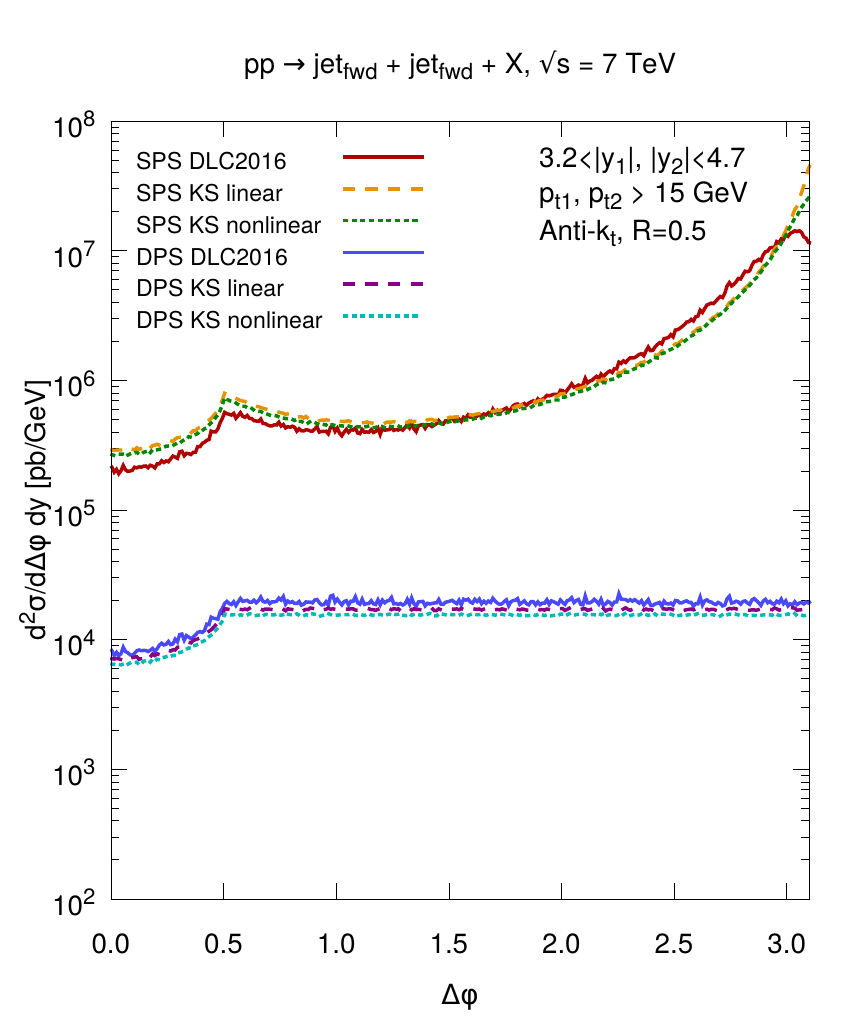}
  \includegraphics[width=0.48\textwidth]{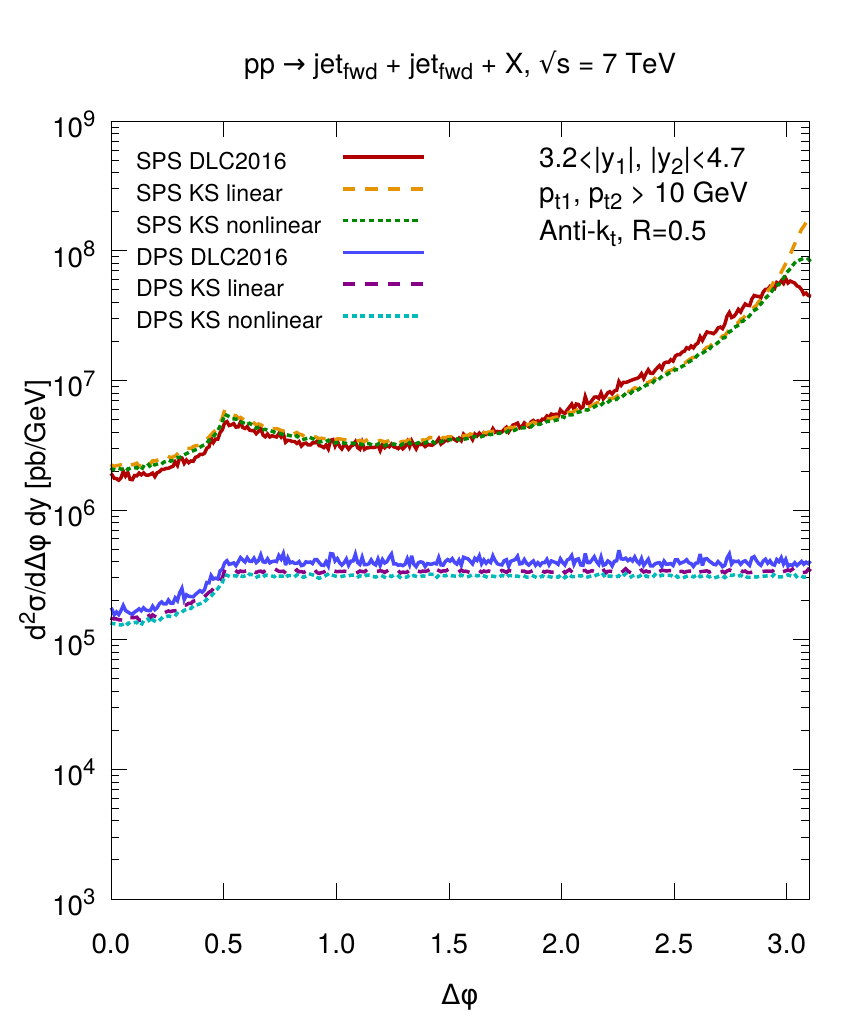}
  \hfill
  \includegraphics[width=0.48\textwidth]{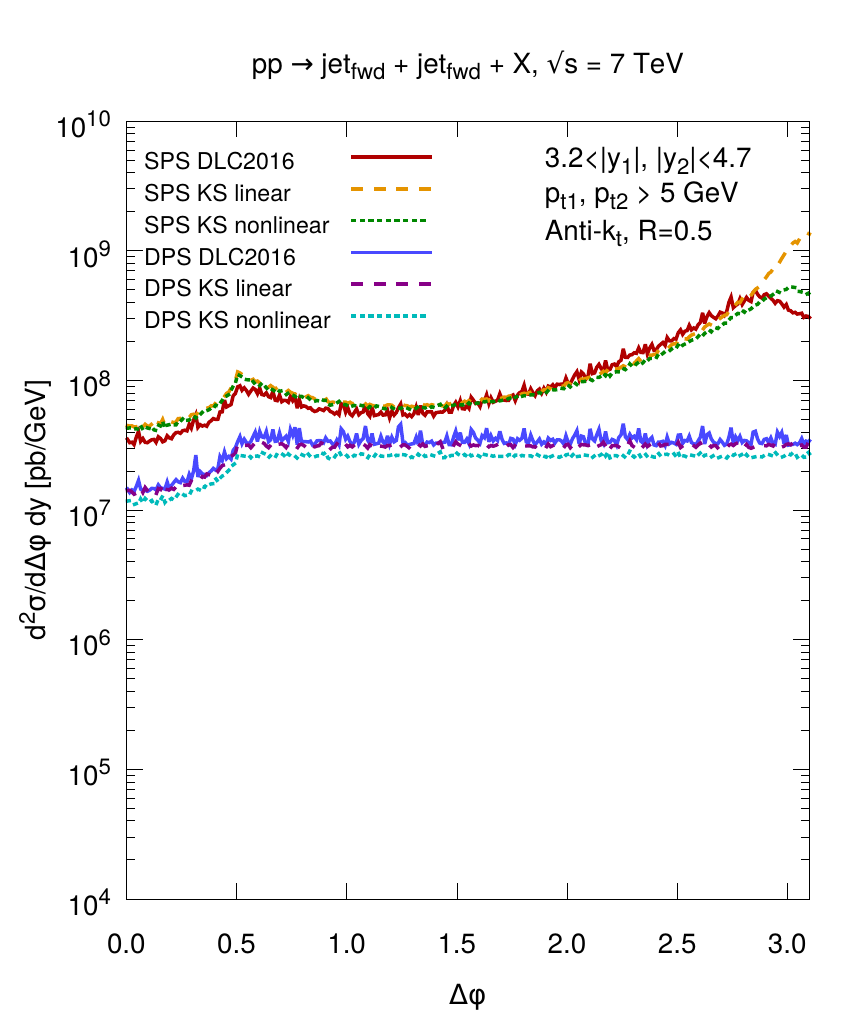}
  \caption{\small{SPS and DPS contribution to forward dijet production for
  various cuts on transverse momentum of a jet at the LHC at $\sqrt{s}=7$ TeV.}
  }
  \label{fig:sps-dps-relative7tev}
\end{figure}

The formula for the SPS contribution to forward dijet cross section reads
\cite{Deak:2009xt,Kutak:2012rf}  
\begin{equation}
  \frac{d\sigma^{pA\rightarrow {\rm dijets}+X}_\text{SPS}}
       {dy_1dy_2dp_{1t}dp_{2t}d\Delta\phi} 
  =
  \frac{p_{1t}p_{2t}}{8\pi^2 (x_1x_2 s)^2}
  \sum_{a,c,d} 
  x_1 f_{a/p}(x_1,\mu^2)\,
  |\overline{{\cal M}_{ag^*\to cd}}|^2
  {\cal F}_{g/A}(x_2,k_t^2)\frac{1}{1+\delta_{cd}}\,,
  \label{eq:hef-formula}
\end{equation}
where
\begin{equation}
x_1  = \frac{1}{\sqrt{s}} \left(|p_{1t}|
e^{y_1}+|p_{2t}| e^{y_2}\right)\,,
\qquad
x_2  = \frac{1}{\sqrt{s}} \left(|p_{1t}|
e^{-y_1}+|p_{2t}| e^{-y_2}\right)\,,
\end{equation}
are the fractions of incoming particles' momenta carried by the partons
participating in the hard interaction 
and
\begin{equation}
  k_{t}^2 = |\mathbold{p_{1t}}+\mathbold{p_{2t}}|^2 = 
  p_{1t}^2 + p_{2t}^2 + 2p_{1t} p_{2t} \cos\Delta\phi\,
  \label{eq:ktglue}
\end{equation}
is an imbalance of the transverse momentum of the two leading jets, which, in
the HEF formalism is equal to the off-shellness of the incoming gluon.
The two leading jets are separated in the transverse plane by the angle
$\Delta\phi$.
 
The expressions for the matrix elements can be found in
Refs.~\cite{Deak:2009xt,Kutak:2012rf}, while the parton densities at the
approximation we are working with are of the same kind as for the single
inclusive jet production.  
Our aim is now to identify and quantify potential corrections to the HEF
framework encapsulated in Eq.~(\ref{eq:hef-formula}).

One of the above may come from double parton scattering \cite{Gaunt:2011xd,Bansal:2014paa,vanHameren:2014ava,Maciula:2014pla,Maciula:2015vza,Astalos:2015ivw,Kutak:2016mik}.  
In general, the cross section for
DPS involves parton density functions which take into account
correlations of partons inside the hadrons before the hard scattering\cite{Diehl:2011yj,Blok:2013bpa,Blok:2012mw,Blok:2011bu}.
However, recent study of Ref.~\cite{Golec-Biernat:2015aza} shows that a
factorized assumption for DPS is largely valid at high scales ($Q^2 >
10^2$~GeV$^2$). Following this observation, we can therefore write

\be
  \frac{d\sigma^{pA\rightarrow {\rm dijets}+X}_\text{DPS}}
       {dy_1d^2p_{1t}dy_2d^2p_{2t}}
  =\frac{1}{\sigma_{\rm
  effective}}\frac{d\sigma}{dy_1d^2p_{1t}}\frac{d\sigma}{dy_2d^2p_{2t}}\,,
\label{eq:dps-approx}
\ee
where $\sigma_{\rm effective}=15\, {\rm mb}$, based on the recent measurement of
the LHCb~\cite{Aaij:2011yc,Aaij:2012dz} collaboration, which confirmed previous
results of D0~\cite{Abazov:2009gc} and CDF~\cite{Abe:1997xk}.
The explicit expressions for DPS cross cross sections are given in
Appendix~\ref{app:dps-formulae}.

The DPS contributions are in general expected to be strong in the low-$p_t$
region of phase space. In order to quantify the role of DPS 
in forward--forward dijet production, we have
calculated the DPS contribution to the azimuthal-angle dependence.
Of course, we expect that, in the approximation
of Eq.~(\ref{eq:dps-approx}), where the correlations between incoming partons
from different pairs are neglected, the contribution will be just of pedestal
type, thus only changing the overall normalization.

\begin{figure}[p]
\centering
  \includegraphics[width=0.48\textwidth]{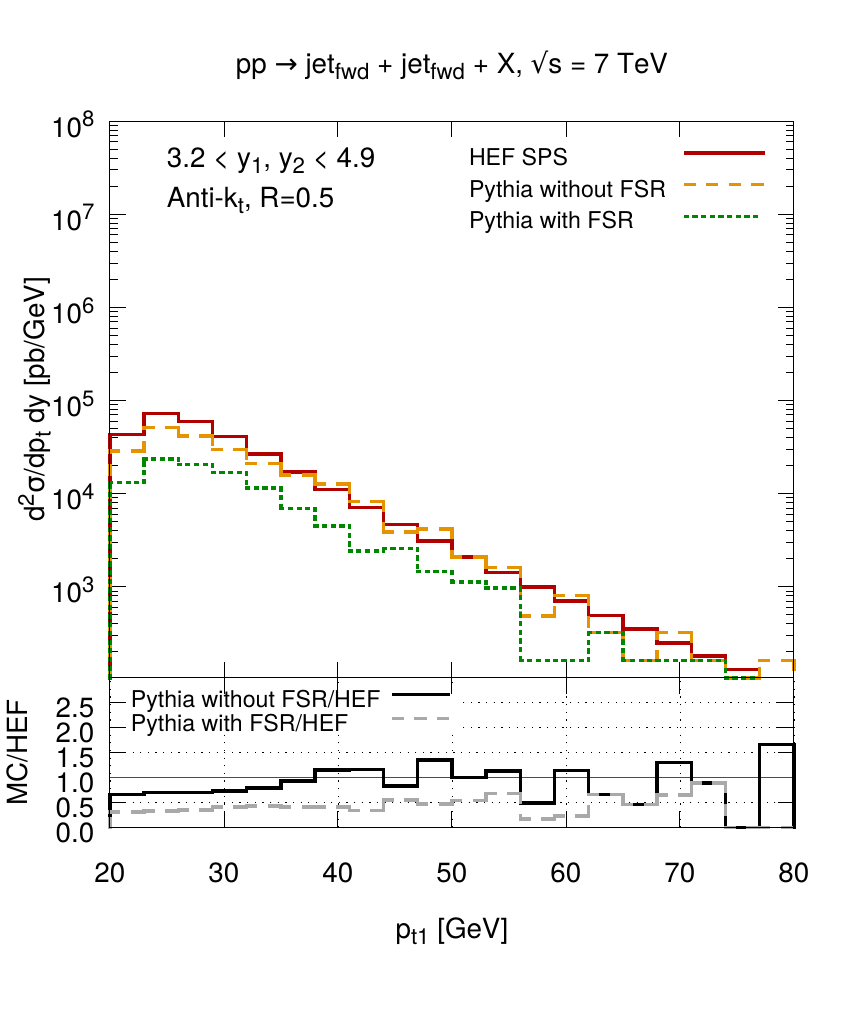}
  \hfill
  \includegraphics[width=0.48\textwidth]{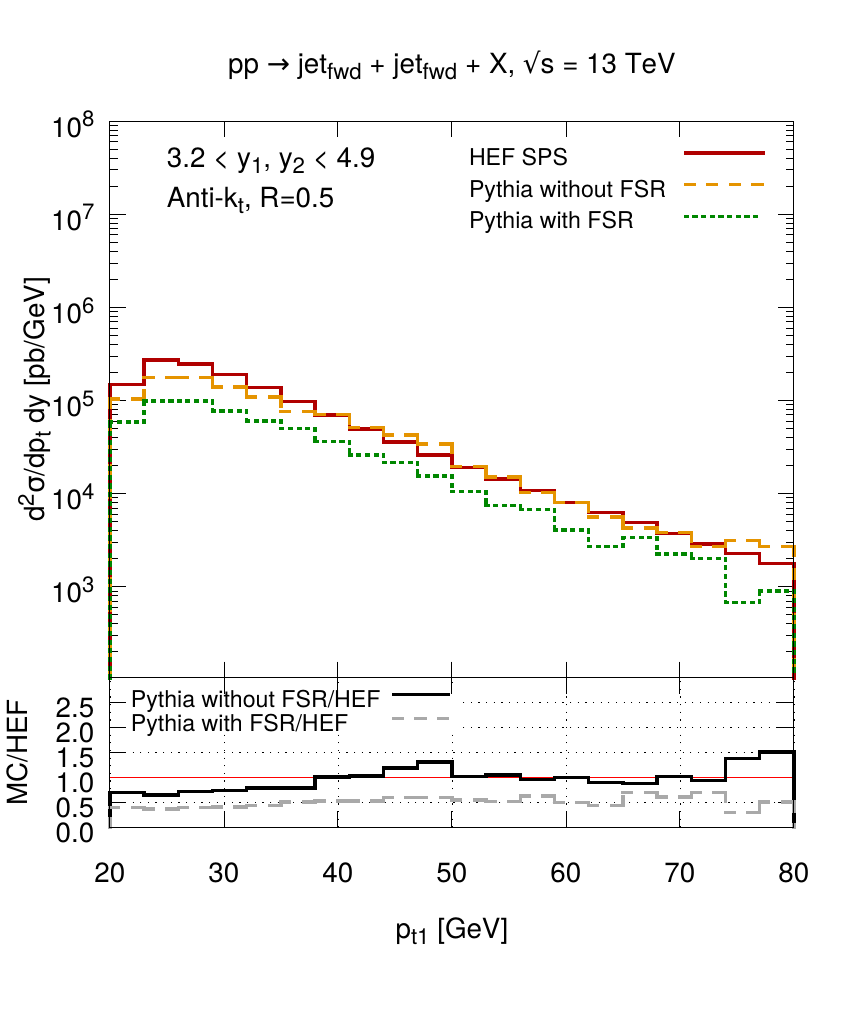}
  \includegraphics[width=0.48\textwidth]{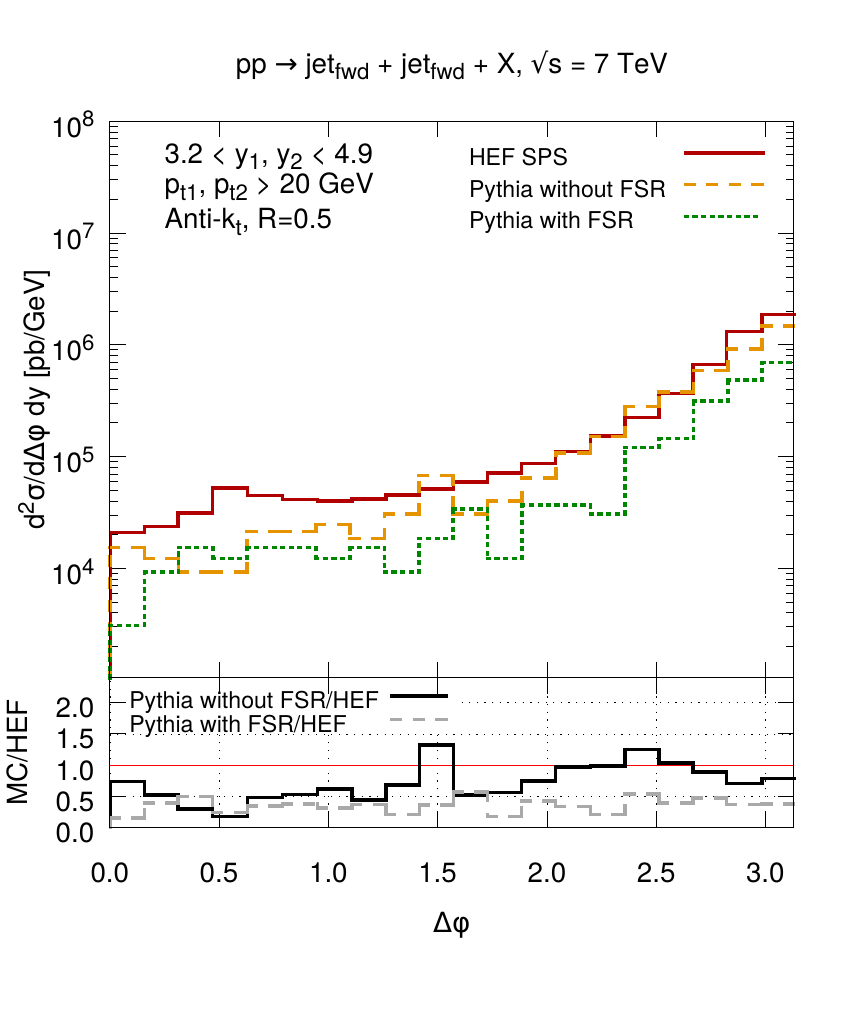}
  \hfill
  \includegraphics[width=0.48\textwidth]{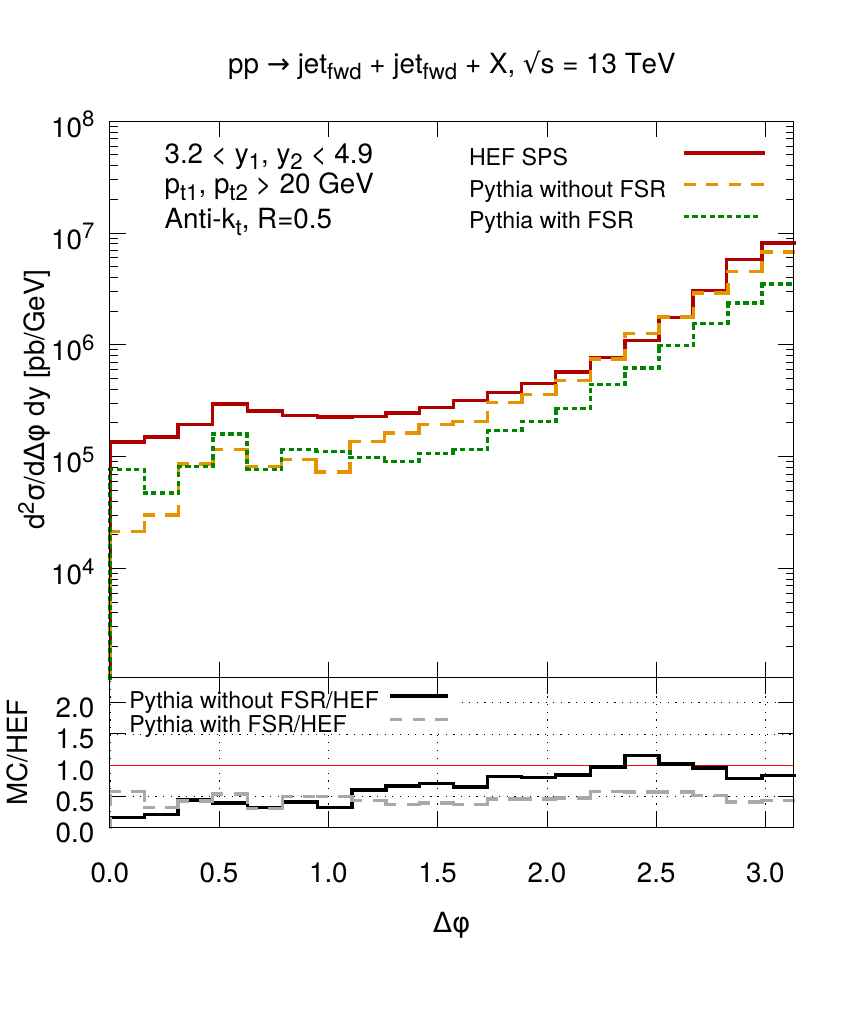}
  \caption{\small{
  Transverse momentum distribution and azimuthal decorrelation in forward dijet
  production. Comparison of predictions from high energy factorization,
  Eq.~(\ref{eq:hef-formula}) with DLC2016 unintegrated
  gluon~\cite{Kutak:2016mik}, and \pythia MC generator.  We checked that
  including MPI has negligible effect on these distributions.
  }}
  \label{fig:hef-vs-pythia}
\end{figure}

\begin{figure}[t]
\centering
  \includegraphics[width=0.42\textwidth]{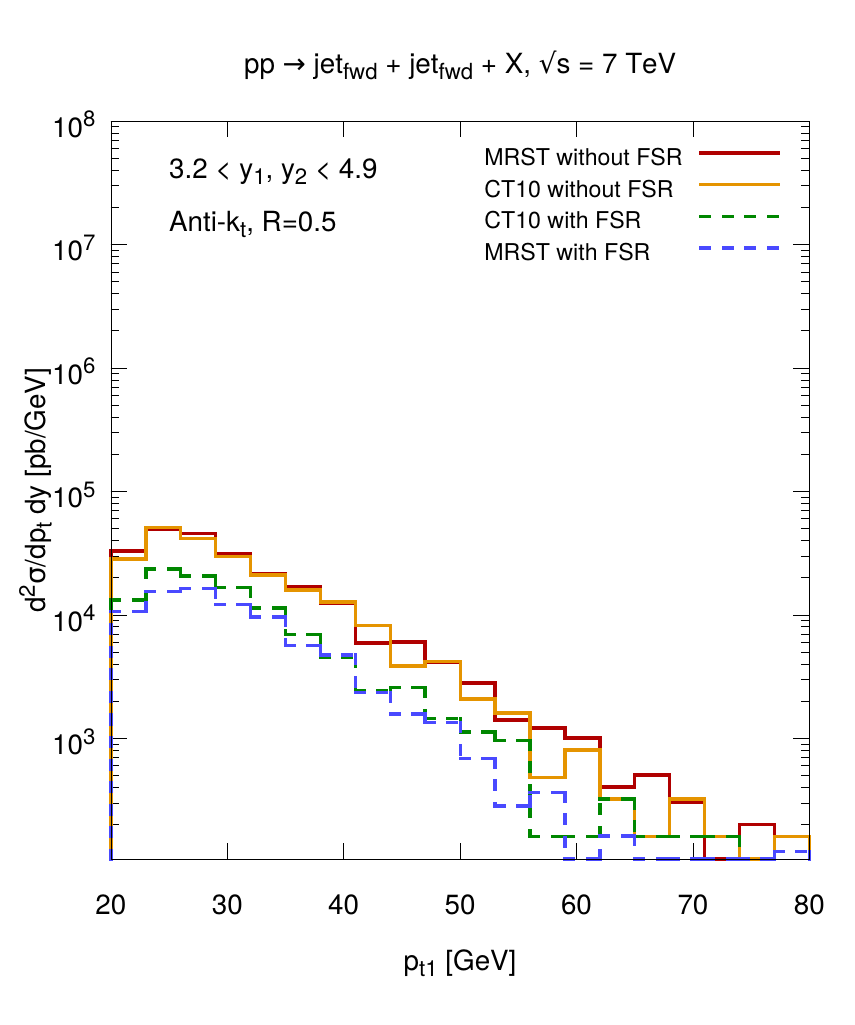}
  \hspace{30pt}
  \includegraphics[width=0.42\textwidth]{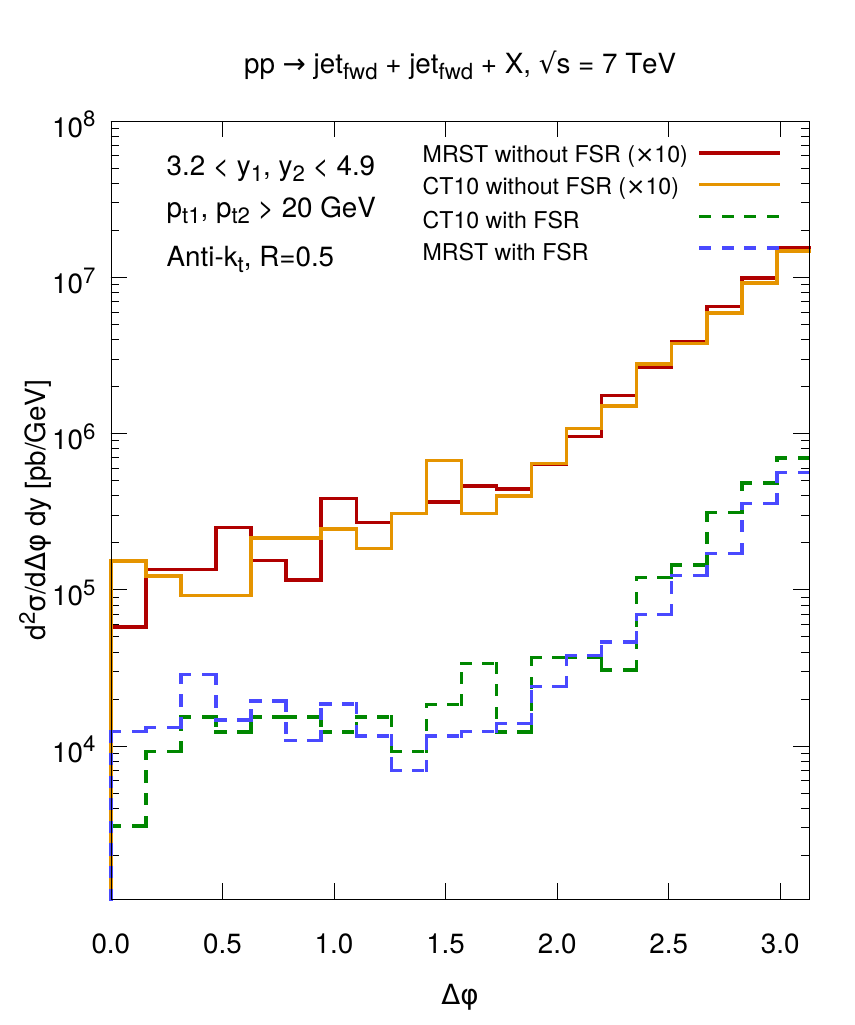}
  \caption{
  \small{Comparison of results for forward dijet spectra obtained with different
  PDFs used in \pythia MC generator: MRST NLO~\cite{Martin:2006qz} and CT10
  NLO~\cite{Lai:2010vv}.
  }}
  \label{fig:pythia-pdfscomparison}
\end{figure}

In Fig. \ref{fig:sps-dps-relative7tev} we show the SPS and the DPS contributions
to the azimuthal angle distributions for various cuts on the
the hardest jet's transverse momentum, set respectively at
35, 15, 10 and 5 GeV. 
We see that the relative contribution of DPS increases with lowering the
transverse momentum jet cut, but it is significantly smaller than SPS at the
experimentally relevant value of 35 GeV.
We have checked that the picture looks very similar at 13 TeV.

\subsection{HEF vs. collinear factorization}
\label{sec:pythia}

We now turn to the comparison of our HEF predictions for the forward dijet
production with the results obtained within collinear factorization \footnote{The preliminary estimate of the result in HEF at 7 TeV has been performed in \cite{Ducloue:2015jba}.}.
To produce the latter, we have used the {\sc
Pythia}~8.2 MC generator. 
To cluster parton level particles into jets, the anti-$k_t$
algorithm~\cite{Cacciari:2008gp} was used, as implemented in the {\sc FastJet}
package~\cite{Cacciari:2011ma}. The cuts on the jets' momenta were set to $p_{t,
1,2}>20\;$GeV and the rapidity window $3.2<y_1,\,y_2<4.9$ was imposed to select
the jets. The jet radius was set to $R=0.5$, as in 
the HEF calculation discussed in preceding subsection.

{\sc Pythia} was set up to generate events including the leading partonic
sub-processes. For the comparison with HEF, the CT10 NLO PDFs~\cite{Lai:2010vv}
were used. 
Runs were performed at two proton--proton collision energies: $7$ and
$13$~TeV. For each energy, two sets of MC data were produced,
distinguished by the final state radiation (FSR) option turned on or off.

The comparison between the HEF and the collinear factorization results in shown
in Fig.~\ref{fig:hef-vs-pythia}. We see that, in the case in which FSR radiation
is turned off, the two formalisms agree quite well in description of the $p_t$
spectra in the whole range of considered values. 
In the case of the azimuthal angle distributions, shown in the lower panel of
Fig.~\ref{fig:hef-vs-pythia}, the results agree in the region of large and moderate angles and differ
in the region of small angles. 
 
We attribute the latter to the effect of different treatment of
singularities in the two frameworks. In HEF, matrix element diverges as the two
outgoing partons become collinear, see Ref.~\cite{Kotko:2015ura}. This
divergence is regularized by the jet algorithm, which is responsible for the
kink around $\Delta\phi=0.5$, seen in the lower panel of
Fig.~\ref{fig:hef-vs-pythia}. 
 On the other hand, in the case of \pythia, the
shape of the distribution is a result of initial state radiation generated via parton shower
matched with the collinear matrix element. Since the collinear matrix elements have different singularity structure than the HEF matrix elements
this leads to different results at small $\Delta\phi$
\cite{Deak:2009xt,Kotko:2015ura}.


In Fig. \ref{fig:pythia-pdfscomparison} we show a comparison between
distributions obtained with \pythia but using two different PDF sets. As we see,
both sets give similar results, hence, the qualitative differences between
\pythia and HEF, seen in Fig.~\ref{fig:hef-vs-pythia}, cannot be attributed to a
choice of PDFs.

In Fig.~\ref{fig:hef-vs-pythia}, we also observe that
turning on the FSR in {\sc Pythia} leads to change in normalization of both the
$p_t$ and $\Delta\phi$ distributions.
The spectra decrease by factor $\sim 2$ for moderate and large $p_t$, as well as
$\Delta\phi$, values. Low-$p_t$ and low-$\Delta\phi$ parts of the distributions
are almost not affected by FSR.
 
The observed difference in normalization of the transverse momentum spectra can
be explained by the energy loss of the leading hard parton that happens readily
via FSR parton shower emissions. 
For a significant fraction of events, this leads to the situation in which the
parton originating from the hard collision splits into two partons separated by
an angle sufficient to produce two lower-$p_t$ jets.
This mechanism takes the high-$p_t$ events from the tail of the spectrum without
FSR and moves them to the region below the jet cut. Hence, they effectively do
not contribute to the observables shown in Fig.~\ref{fig:hef-vs-pythia}.

Finally, we mention that we have checked explicitly that the picture of
Fig.~\ref{fig:hef-vs-pythia} persists if \pythia events are supplemented with
multi-parton interactions~(MPI). Hence,  forward dijet production in the
collinear factorization framework is weakly sensitive to MPIs, which is
consistent with the negligible effect of DPS in HEF, which we demonstrated in
Section~\ref{sec:single-inclusive}.

\section{Conclusions}

In this letter, we have studied double and single inclusive forward jet
production at the LHC within two formalisms: the high energy factorisation~(HEF)
and the collinear factorization. 
 
We have demonstrated that the HEF framework describes well the single inclusive
jet production at the LHC, at the center-of-mass energies of 7 and 13~TeV, and
the main uncertainty comes from the unintegrated parton distributions.  In this
context, we have observed that the contribution from the off-shell quarks is
negligible for forward jet production. 

We have also explicitly shown that, for typical experimental cuts
used in inclusive dijet production processes, the double parton
scattering effects can be safely neglected.
Finally, our study shows that the effect of the final state radiation
is not negligible and it leads to change of normalization of differential
distributions in forward dijet production.

\section*{Acknowledgments}

We are grateful to Andreas van Hameren for careful reading of the manuscript
and numerous useful suggestions.
We also appreciate comments from Piotr Kotko and Hannes Jung.
The work of M.B. and K.K. has been supported by Narodowe Centrum Nauki~(NCN)
with Sonata Bis grant DEC-2013/10/E/ST2/00656. M.D. acknowledges
hospitality of Instytut Fizyki Jadrowej, where the project was initiated, as
well as the support from the NCN grant DEC-2013/10/E/ST2/00656.


\appendix

\section{Matrix elements}
\label{app:single-inclusive-matrix-elements}

The matrix elements squared for three-parton processes, used in the calculation
of the single inclusive forward jet distributions in
Section~\ref{sec:single-inclusive} read
\begin{itemize}
\item
$gg^*\to g$
\begin{equation}
\overline{|{\cal M}_{gg^*\to g}|}^2=
4 g_s^2\,\frac{C_A}{N_c^2-1}\frac{(k\cdot q)^2}{k_t^2}\,,
\end{equation}
\item
$qg^*\to q$
\begin{equation}
\overline{|{\cal M}_{qg^*\to g}|}^2=
4 g_s^2\,\frac{C_F}{N_c^2-1}\,\frac{(k\cdot q)^2}{k_t^2}\,,
\end{equation}
\item
$gq^*\to q$
\begin{equation}
\overline{|{\cal M}_{gq^*\to g}|}^2=
g_s^2 \frac{C_F}{N_c^2-1}(k\cdot q)\,,
\end{equation}
\item
$\bar{q}q^*\to g$
\begin{equation}
\overline{|{\cal M}_{\bar{q}q^*\to g}|}^2=
g_s^2 \frac{C_F}{N_c}(k\cdot q)\,.
\end{equation}
\end{itemize}
In the above, $k$ and $q$ are the momenta of the off-shell and on-shell partons,
respectively. $k_t$ is the transverse component of the off-shell momentum, $g_s$
is a strong coupling and  $C_i$ is a colour factor of the emitter: $C_F$ for a
quark and $C_A$ for a gluon.

\section{Double-parton scattering formulae}
\label{app:dps-formulae}

The explicit expression for the DPS contribution in the factorized
approximation reads
\be
\begin{aligned}
&\frac{d\sigma_{\rm DPS}}{dy_1dy_2dp_{1t}dp_{2t}d\Delta\phi} = \frac{1}{\sigma_{\scriptscriptstyle{eff}}}\frac{\pi}{8}\frac{p_{1t}}{(x_1x_2s)^2} \frac{p_{2t}}{(\bar x_1 \bar x_2s)^2}\\
&\qquad \times \bigg( \overline{|{\cal M}_{\scriptscriptstyle{gg^{*}\to g}}|}^2 x_1f_{g/_A}(x_1)+ \sum_{i=1}^{n_f}\overline{|{\cal M}_{\scriptscriptstyle{qg^{*}\to q}}|}^2 x_1f_{{q(i)}/_A}(x_1)  \bigg)\\
&\qquad \qquad \times \bigg( \overline{|{\bar{\cal M}}_{\scriptscriptstyle{gg^{*}\to g}}|}^2 \bar x_1\bar f_{g/_A}(\bar x_1)+ \sum_{i=1}^{n_f}\overline{|{\bar{\cal M}}_{\scriptscriptstyle{qg^{*}\to q}}|}^2 \bar x_1\bar f_{{q(i)}/_A}(\bar x_1)  \bigg)\\
&\qquad \qquad \qquad \times {\cal F}_{{g^{*}}/_B}(x_2,p_{1t}^2)\bar
{\cal F}_{{g^{*}}/_B}(\bar x_2,p_{2t}^{2})\theta(1-x_1-\bar x_1) \theta(1-x_2-\bar
x_2) \,,
\end{aligned}
\ee
where, in order to be compatible with the SPS formula (\ref{eq:hef-formula}) we introduced an
auxiliary azimuthal angle between the final state jets, $\Delta\phi$. 
The notation follows that of Eq.~(\ref{eq:hef-formula} except that now, each of
the incoming particles provides a pair of partons, whose energy
fractions are given by $x_1$ and $\bar x_1$ for hadron $A$, and
$x_2$ and $\bar x_2$ for hadron $B$. The theta functions guarantee that a pair
of partons from a single hadron does not carry more than 100\% of the hadron's
energy.

\begin{small}

\end{small}

\begin{thebibliography}{99}

\bibitem{Kuraev:1977fs}
  E.~A.~Kuraev, L.~N.~Lipatov and V.~S.~Fadin,
  Sov.\ Phys.\ JETP {\bf 45} (1977) 199
   [Zh.\ Eksp.\ Teor.\ Fiz.\  {\bf 72} (1977) 377].

\bibitem{Balitsky:1978ic}
  I.~I.~Balitsky and L.~N.~Lipatov,
  Sov.\ J.\ Nucl.\ Phys.\  {\bf 28} (1978) 822
   [Yad.\ Fiz.\  {\bf 28} (1978) 1597].

\bibitem{Kuraev:1976ge}
  E.~A.~Kuraev, L.~N.~Lipatov and V.~S.~Fadin,
  Sov.\ Phys.\ JETP {\bf 44} (1976) 443
   [Zh.\ Eksp.\ Teor.\ Fiz.\  {\bf 71} (1976) 840].

\bibitem{Balitsky:1995ub}
  I.~Balitsky,
  Nucl.\ Phys.\  {\bf B463 } (1996)  99-160.

\bibitem{JalilianMarian:1997dw}
  J.~Jalilian-Marian, A.~Kovner and H.~Weigert,
  Phys.\ Rev.\ D {\bf 59} (1998) 014015
  doi:10.1103/PhysRevD.59.014015
  [hep-ph/9709432].

\bibitem{Kovchegov:1999yj}
  Y.~V.~Kovchegov,
  Phys.\ Rev.\  D {\bf 60} (1999) 034008.

\bibitem{Kovchegov:1999ua}
  Y.~V.~Kovchegov,
  Phys.\ Rev.\  D {\bf 61} (2000) 074018.

\bibitem{Kovner:2005nq}
  A.~Kovner and M.~Lublinsky,
  Phys.\ Rev.\ D {\bf 71} (2005) 085004
  doi:10.1103/PhysRevD.71.085004
  [hep-ph/0501198].

\bibitem{Kutak:2011fu}
  K.~Kutak, K.~Golec-Biernat, S.~Jadach and M.~Skrzypek,
  JHEP {\bf 1202} (2012) 117
  doi:10.1007/JHEP02(2012)117
  [arXiv:1111.6928 [hep-ph]].

\bibitem{Catani:1990eg}
  S.~Catani, M.~Ciafaloni and F.~Hautmann,
  Nucl.\ Phys.\ B {\bf 366} (1991) 135.

\bibitem{Kotko:2015ura}
  P.~Kotko, K.~Kutak, C.~Marquet, E.~Petreska, S.~Sapeta and A.~van Hameren,
  JHEP {\bf 1509} (2015) 106
  doi:10.1007/JHEP09(2015)106
  [arXiv:1503.03421 [hep-ph]].

\bibitem{Rothstein:2016bsq}
  I.~Z.~Rothstein and I.~W.~Stewart,
  arXiv:1601.04695 [hep-ph].

\bibitem{Sjostrand:2014zea}
  T.~Sjöstrand {\it et al.},
  Comput.\ Phys.\ Commun.\  {\bf 191} (2015) 159
  doi:10.1016/j.cpc.2015.01.024
  [arXiv:1410.3012 [hep-ph]].

\bibitem{Bellm:2015jjp}
  J.~Bellm {\it et al.},
  arXiv:1512.01178 [hep-ph].


\bibitem{Sapeta:2015gee}
  S.~Sapeta,
  arXiv:1511.09336 [hep-ph].

\bibitem{Bury:2015dla}
  M.~Bury and A.~van Hameren,
  Comput.\ Phys.\ Commun.\  {\bf 196} (2015) 592
  doi:10.1016/j.cpc.2015.06.023
  [arXiv:1503.08612 [hep-ph]].

\bibitem{Kotko_LxJet}
P.~Kotko, {\it {LxJet}},  2013.
\newblock {C++} Monte Carlo program.

\bibitem{Jung:2010si}
  H.~Jung {\it et al.},
  Eur.\ Phys.\ J.\ C {\bf 70} (2010) 1237
  doi:10.1140/epjc/s10052-010-1507-z
  [arXiv:1008.0152 [hep-ph]].


\bibitem{vanHameren:2014lna}
  A.~van Hameren, P.~Kotko, K.~Kutak, C.~Marquet and S.~Sapeta,
  Phys.\ Rev.\ D {\bf 89} (2014) no.9,  094014
  doi:10.1103/PhysRevD.89.094014
  [arXiv:1402.5065 [hep-ph]].


\bibitem{vanHameren:2014ala}
  A.~van Hameren, P.~Kotko, K.~Kutak and S.~Sapeta,
  Phys.\ Lett.\ B {\bf 737} (2014) 335
  doi:10.1016/j.physletb.2014.09.005
  [arXiv:1404.6204 [hep-ph]].

\bibitem{Chirilli:2011km}
  G.~A.~Chirilli, B.~W.~Xiao and F.~Yuan,
  Phys.\ Rev.\ Lett.\  {\bf 108} (2012) 122301
  doi:10.1103/PhysRevLett.108.122301
  [arXiv:1112.1061 [hep-ph]].

\bibitem{Altinoluk:2014eka}
  T.~Altinoluk, N.~Armesto, G.~Beuf, A.~Kovner and M.~Lublinsky,
  Phys.\ Rev.\ D {\bf 91} (2015) no.9,  094016
  doi:10.1103/PhysRevD.91.094016
  [arXiv:1411.2869 [hep-ph]].

\bibitem{Dumitru:2005gt}
  A.~Dumitru, A.~Hayashigaki and J.~Jalilian-Marian,
  Nucl.\ Phys.\ A {\bf 765} (2006) 464
  doi:10.1016/j.nuclphysa.2005.11.014
  [hep-ph/0506308].


\bibitem{Kimber:1999xc}
  M.~A.~Kimber, A.~D.~Martin and M.~G.~Ryskin,
  Eur.\ Phys.\ J.\ C {\bf 12} (2000) 655
  doi:10.1007/s100520000326
  [hep-ph/9911379].

\bibitem{Kimber:2001sc}
  M.~A.~Kimber, A.~D.~Martin and M.~G.~Ryskin,
  Phys.\ Rev.\ D {\bf 63} (2001) 114027
  doi:10.1103/PhysRevD.63.114027
  [hep-ph/0101348].

\bibitem{Lublinsky:2001yi}
  M.~Lublinsky, E.~Gotsman, E.~Levin and U.~Maor,
  Nucl.\ Phys.\ A {\bf 696} (2001) 851
  doi:10.1016/S0375-9474(01)01150-2
  [hep-ph/0102321].

\bibitem{GolecBiernat:2001if}
  K.~J.~Golec-Biernat, L.~Motyka and A.~M.~Stasto,
  Phys.\ Rev.\ D {\bf 65} (2002) 074037
  doi:10.1103/PhysRevD.65.074037
  [hep-ph/0110325].

\bibitem{Kutak:2003bd}
  K.~Kutak and J.~Kwiecinski,
  Eur.\ Phys.\ J.\  C {\bf 29} (2003) 521
  [arXiv:hep-ph/0303209].

\bibitem{Kutak:2004ym}
  K.~Kutak and A.~M.~Stasto,
  Eur.\ Phys.\ J.\ C {\bf 41} (2005) 343
  doi:10.1140/epjc/s2005-02223-0
  [hep-ph/0408117].

\bibitem{Albacete:2010sy}
  J.~L.~Albacete, N.~Armesto, J.~G.~Milhano, P.~Quiroga-Arias and C.~A.~Salgado,
  Eur.\ Phys.\ J.\ C {\bf 71} (2011) 1705
  doi:10.1140/epjc/s10052-011-1705-3
  [arXiv:1012.4408 [hep-ph]].

\bibitem{Lappi:2015fma}
  T.~Lappi and H.~Mäntysaari,
  Phys.\ Rev.\ D {\bf 91} (2015) 7,  074016
  doi:10.1103/PhysRevD.91.074016
  [arXiv:1502.02400 [hep-ph]].

\bibitem{vanHameren:2012uj}
  A.~van Hameren, P.~Kotko and K.~Kutak,
  JHEP {\bf 1212} (2012) 029
  doi:10.1007/JHEP12(2012)029
  [arXiv:1207.3332 [hep-ph]].
  
\bibitem{vanHameren:2012if}
  A.~van Hameren, P.~Kotko and K.~Kutak,
  JHEP {\bf 1301} (2013) 078
  doi:10.1007/JHEP01(2013)078
  [arXiv:1211.0961 [hep-ph]].
  
\bibitem{vanHameren:2013csa}
  A.~van Hameren, K.~Kutak and T.~Salwa,
  Phys.\ Lett.\ B {\bf 727} (2013) 226
  doi:10.1016/j.physletb.2013.10.039
  [arXiv:1308.2861 [hep-ph]].

\bibitem{Nefedov:2013ywa}
  M.~A.~Nefedov, V.~A.~Saleev and A.~V.~Shipilova,
  Phys.\ Rev.\ D {\bf 87} (2013) 9,  094030
  doi:10.1103/PhysRevD.87.094030
  [arXiv:1304.3549 [hep-ph]].

\bibitem{Chatrchyan:2012gwa}
  S.~Chatrchyan {\it et al.} [CMS Collaboration],
  JHEP {\bf 1206} (2012) 036
  [arXiv:1202.0704 [hep-ex]].

\bibitem{CMS13}
CMS-PAS-SMP-15-007. Measurement of the double-differential inclusive jet cross section at $\sqrt{s}=13$ TeV

\bibitem{Lai:2010vv}
  H.~L.~Lai, M.~Guzzi, J.~Huston, Z.~Li, P.~M.~Nadolsky, J.~Pumplin and C.-P.~Yuan,
  Phys.\ Rev.\ D {\bf 82} (2010) 074024
  doi:10.1103/PhysRevD.82.074024
  [arXiv:1007.2241 [hep-ph]].

\bibitem{Kutak:2012rf}
  K.~Kutak and S.~Sapeta,
  Phys.\ Rev.\ D {\bf 86} (2012) 094043
  doi:10.1103/PhysRevD.86.094043
  [arXiv:1205.5035 [hep-ph]].

\bibitem{Kwiecinski:1997ee}
  J.~Kwiecinski, A.~D.~Martin and A.~M.~Stasto,
  Phys.\ Rev.\ D {\bf 56} (1997) 3991
  doi:10.1103/PhysRevD.56.3991
  [hep-ph/9703445].

\bibitem{Kutak:2014wga}
  K.~Kutak,
  Phys.\ Rev.\ D {\bf 91} (2015) no.3,  034021
  doi:10.1103/PhysRevD.91.034021
  [arXiv:1409.3822 [hep-ph]].

\bibitem{Kutak:2016mik}
  K.~Kutak, R.~Maciula, M.~Serino, A.~Szczurek and A.~van Hameren,
  arXiv:1602.06814 [hep-ph].


\bibitem{Ducloue:2015jba}
  B.~Ducloué, L.~Szymanowski and S.~Wallon,
  Phys.\ Rev.\ D {\bf 92} (2015) no.7,  076002
  doi:10.1103/PhysRevD.92.076002
  [arXiv:1507.04735 [hep-ph]].


\bibitem{forward}
  S.~Sapeta and M.~Bury, \forward, code on request

\bibitem{Dominguez:2011wm}
  F.~Dominguez, C.~Marquet, B.~W.~Xiao and F.~Yuan,
  Phys.\ Rev.\ D {\bf 83} (2011) 105005
  doi:10.1103/PhysRevD.83.105005
  [arXiv:1101.0715 [hep-ph]].

\bibitem{Deak:2009xt}
  M.~Deak, F.~Hautmann, H.~Jung and K.~Kutak,
  JHEP {\bf 0909} (2009) 121
  doi:10.1088/1126-6708/2009/09/121
  [arXiv:0908.0538 [hep-ph]].

\bibitem{Gaunt:2011xd}
  J.~R.~Gaunt and W.~J.~Stirling,
  JHEP {\bf 1106} (2011) 048
  doi:10.1007/JHEP06(2011)048
  [arXiv:1103.1888 [hep-ph]].

\bibitem{Bansal:2014paa}
  S.~Bansal {\it et al.},
  arXiv:1410.6664 [hep-ph].

\bibitem{vanHameren:2014ava}
  A.~van Hameren, R.~Maciula and A.~Szczurek,
  Phys.\ Rev.\ D {\bf 89} (2014) no.9,  094019
  doi:10.1103/PhysRevD.89.094019
  [arXiv:1402.6972 [hep-ph]].

\bibitem{Maciula:2014pla}
  R.~Maciula and A.~Szczurek,
  Phys.\ Rev.\ D {\bf 90} (2014) no.1,  014022
  doi:10.1103/PhysRevD.90.014022
  [arXiv:1403.2595 [hep-ph]].

\bibitem{Maciula:2015vza}
  R.~Maciuła and A.~Szczurek,
  Phys.\ Lett.\ B {\bf 749} (2015) 57
  doi:10.1016/j.physletb.2015.07.035
  [arXiv:1503.08022 [hep-ph]].

\bibitem{Astalos:2015ivw}
  R.~Astalos {\it et al.},
  arXiv:1506.05829 [hep-ph].

\bibitem{Diehl:2011yj}
  M.~Diehl, D.~Ostermeier and A.~Schafer,
  JHEP {\bf 1203} (2012) 089
  doi:10.1007/JHEP03(2012)089
  [arXiv:1111.0910 [hep-ph]].


\bibitem{Blok:2013bpa}
  B.~Blok, Y.~Dokshitzer, L.~Frankfurt and M.~Strikman,
  Eur.\ Phys.\ J.\ C {\bf 74} (2014) 2926
  doi:10.1140/epjc/s10052-014-2926-z
  [arXiv:1306.3763 [hep-ph]].


\bibitem{Blok:2012mw}
  B.~Blok, Y.~Dokshitzer, L.~Frankfurt and M.~Strikman,
  arXiv:1206.5594 [hep-ph].

\bibitem{Blok:2011bu}
  B.~Blok, Y.~Dokshitser, L.~Frankfurt and M.~Strikman,
  Eur.\ Phys.\ J.\ C {\bf 72} (2012) 1963
  doi:10.1140/epjc/s10052-012-1963-8
  [arXiv:1106.5533 [hep-ph]].


\bibitem{Golec-Biernat:2015aza}
  K.~Golec-Biernat, E.~Lewandowska, M.~Serino, Z.~Snyder and A.~M.~Stasto,
  Phys.\ Lett.\ B {\bf 750} (2015) 559
  [arXiv:1507.08583 [hep-ph]].

\bibitem{Aaij:2011yc}
  R.~Aaij {\it et al.} [LHCb Collaboration],
  Phys.\ Lett.\ B {\bf 707} (2012) 52
  doi:10.1016/j.physletb.2011.12.015
  [arXiv:1109.0963 [hep-ex]].

\bibitem{Aaij:2012dz}
  R.~Aaij {\it et al.} [LHCb Collaboration],
  JHEP {\bf 1206} (2012) 141
   Addendum: [JHEP {\bf 1403} (2014) 108]
  doi:10.1007/JHEP03(2014)108, 10.1007/JHEP06(2012)141
  [arXiv:1205.0975 [hep-ex]].

\bibitem{Abazov:2009gc}
  V.~M.~Abazov {\it et al.} [D0 Collaboration],
  Phys.\ Rev.\ D {\bf 81} (2010) 052012
  doi:10.1103/PhysRevD.81.052012
  [arXiv:0912.5104 [hep-ex]].

\bibitem{Abe:1997xk}
  F.~Abe {\it et al.} [CDF Collaboration],
  Phys.\ Rev.\ D {\bf 56} (1997) 3811.
  doi:10.1103/PhysRevD.56.3811

\bibitem{Martin:2006qz}
  A.~D.~Martin, W.~J.~Stirling and R.~S.~Thorne,
  Phys.\ Lett.\ B {\bf 636} (2006) 259
  doi:10.1016/j.physletb.2006.03.054
  [hep-ph/0603143].

\bibitem{Cacciari:2008gp}
  M.~Cacciari, G.~P.~Salam and G.~Soyez,
  JHEP {\bf 0804} (2008) 063
  doi:10.1088/1126-6708/2008/04/063
  [arXiv:0802.1189 [hep-ph]].

\bibitem{Cacciari:2011ma}
  M.~Cacciari, G.~P.~Salam and G.~Soyez,
  Eur.\ Phys.\ J.\ C {\bf 72} (2012) 1896
  doi:10.1140/epjc/s10052-012-1896-2
  [arXiv:1111.6097 [hep-ph]].




\end{thebibliography}
\end{document}